\documentstyle[aps,epsfig,multicol]{revtex}
\newcommand{\be}{\begin{equation}}
\newcommand{\ee}{\end{equation}}
\newcommand{\bea}{\begin{eqnarray}}
\newcommand{\eea}{\end{eqnarray}}
\newcommand{\HH}{{\cal H}}

\newcommand{\ra}{\rangle}

\begin{document}
\title{Landau-Zener transitions in a multilevel system. An exact result.}
\author{A.\,V. Shytov}
\address{Lyman Laboratory, Physics Department, Harvard University, 
    1 Oxford street, Cambridge, MA02138}
\address{L.\,D. Landau Institute for Theoretical Physics,
    Kosygin street, 2, Moscow 117934, Russia}

\maketitle
\begin{abstract}
We study the S-matrix for the transitions at an avoided crossing 
of several energy levels,
which is a multilevel generalization of the Landau-Zener problem.
We demonstrate that, by extending the Schr\"odinger evolution
to complex time, one can obtain an exact answer for some of the transition
amplitudes. 
Similar to the Landau-Zener case, our result covers both
the adiabatic regime (slow evolution.) and the diabatic regime 
(fast evolution). 
The form of the exact transition amplitude coincides with that obtained
in a sequential pairwise level crossing approximation,
in accord with the conjecture of Brundobler and Elser~\cite{Elser93}.
%% Furthermore,
\end{abstract}

\begin{multicols}{2}

\narrowtext
Landau-Zener (LZ) transitions~\cite{Landau32,Zener32} 
between two energy states at an avoided 
level crossing is one of the few exactly solvable problems of time-dependent
quantum evolution. The LZ theory has found many applications,
and is used to describe a variety of physical systems, 
ranging from atomic and molecular physics~\cite{Nakamura_book}, 
to mesoscopic physics~\cite{Wilkinson93}. 

However, since most of the problems of interest involve more than two 
energy levels, with transitions between several levels happening simultaneously,
it is desirable to extend the LZ theory to multilevel problems. 
A number of interesting generalizations of the LZ problem have been
proposed, in which several energy levels cross at the same time.
In some cases, including the so-called bow-tie 
model~\cite{Ostrovsky97,Demkov00,Demkov01}, 
the high spin model~\cite{Pokrovsky00}, 
one can construct an analytic solution. 
The solutions~\cite{Ostrovsky97,Demkov00,Demkov01,Pokrovsky00}
rely on a special form of the coupling Hamiltonian
or on its symmetry,
which preserves the integrability of the problem.
One of the most efficient methods proposed to treat 
the generalized LZ problems
is the contour integration approach~\cite{Demkov68}.

Much less is known about the general multilevel LZ problem, 
defined by a time-dependent Hamiltonian of the form
\be\label{eq:A+Bt}
\HH(t)=\hat\Delta + \hat\nu t
\ee
were $\hat\Delta$, $\hat\nu$ are hermitian matrices $n\times n$ 
of a general form. In this
problem, one is interested in the evolution 
\be\label{eq:Schroedinger}
i\frac{\partial \psi}{\partial t}
=\HH(t)\psi
\ee 
from large negative 
to large positive times. Since $\hat\nu t$ is the leading term in the  Hamiltonian
at large $t$, it is natural
to analyze the evolution in terms of an S-matrix in the basis of states
that diagonalize $\hat\nu$.

One can construct an approximate solution in the `weak 
coupling' limit, when the off-diagonal
elements of $\hat\Delta$ are weak compared to the diagonal ones,
$|\Delta_{ij}|\ll |\Delta_{ii}|$, ($i \neq j$).
Since in this case the anticrossing gaps of the adiabatic levels
(which sometimes are also called `frozen levels') 
are much smaller than the level separation,
the weak coupling regime can be analyzed in a sequential two-level 
LZ approximation~\cite{Elser93}, 
similar to the energy diffusion problem in 
mesoscopic systems~\cite{Wilkinson93}.

Brundobler and Elser have noticed~\cite{Elser93} that the formula 
for certain diagonal elements of the S-matrix obtained in
such an approximation remains numerically very accurate 
even at strong coupling, when this approximation supposedly
breaks down.
This surprising observation, based on a computer simulation 
of the problem (\ref{eq:Schroedinger}),(\ref{eq:A+Bt}), 
has led Brundobler and Elser to a conjecture that
the sequential LZ approximation can in some cases give
an exact result.

Below we show that this conjecture is indeed correct. We shall use the method
of continuing the Schr\"odinger evolution (\ref{eq:Schroedinger})
to complex time, and matching the $t\to\pm\infty$ asymptotics
of the evolving state $\psi(t)$. Our method gives exact results only 
for some of the transition amplitudes, corresponding to the transitions
$|1\ra_{t=-\infty} \to |1\ra_{t=+\infty}$,
$|n\ra_{t=-\infty} \to |n\ra_{t=+\infty}$.

%%%%%%%%%%%%%%%%%%% figure 1 %%%%%%%%%%%%%%%%%%%%%%%%%%%%%%%%%%%
\begin{figure}[t]
\centerline{
\begin{minipage}[t]{3.5in}
\vspace{-10pt}
\hspace{-10pt}
\centering
\includegraphics[width=3in]{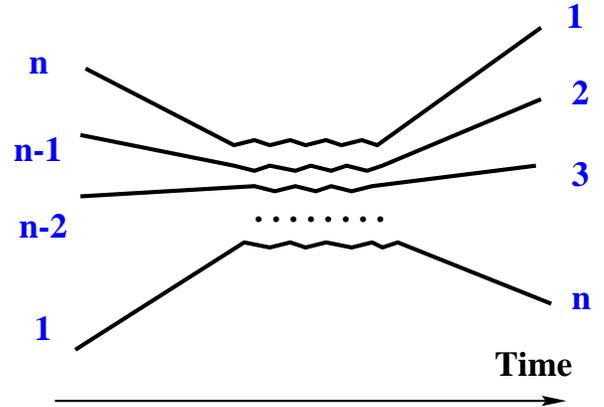}
\end{minipage}
}
\vspace{0.15cm}
\caption[]{
Time evolution of the adiabatic (frozen) energy levels of the 
Hamiltonian (\ref{eq:A+Bt}). The transitions
analyzed in this work are
$|1\ra_{t=-\infty} \to |1\ra_{t=+\infty}$ and
$|n\ra_{t=-\infty} \to |n\ra_{t=+\infty}$.
}
\label{fig:energy evolution}
\end{figure}
%%%%%%%%%%%%%%%%%%%%%%%%%%%%%%%%%%%%%%%%%%%%%%%%%%%%%%%%%%%%%%%%%

It is convenient to analyze the problem in the basis of states that diagonalize
$\hat\nu$:
%% I found that certain transition probabilities for multilevel Landau--Zener
%% (LZ) problem can be found exactly. Consider the Hamiltonian of the 
%% form 
%% \begin{equation}
%% \label{Hamiltonian}
%% \hat{H} = \hat{\nu} t + \hat{\Delta} 
%% \end{equation}
%% acting on the $n$-dimensional Hilbert space:
%% \begin{equation}
%% i \frac{\partial \psi}{\partial t} = \hat{H} \psi \,. 
%% \end{equation}
%% Both $\hat{\nu}$ and $\hat{\Delta}$, are Hermitian.
%% It is convenient to consider the problem in the basis in 
%% which $\nu$ is diagonal:
%
\begin{equation}
\hat{\nu} = 
\left(
  \begin{array}{cccc}
  \nu_{1}                             \\
         & \nu_{2}                    \\
         &         & \ddots           \\ 
	 &         &        & \nu_{n}  
  \end{array}
\right)
\,. 
\end{equation}
This basis is {\it not} the basis 
of adiabatic states, however, for $t \to \pm\infty$ those 
states coincide with the adiabatic states, up to permutation. 
It is also convenient to order the eigenvalues $\nu_i$ so that
\begin{equation}
\nu_1 < \nu_2 <  {\ldots}  < \nu_n \,. 
\end{equation}
For simplicity, we assume that the eigenvalues are non-degenerate: 
$\nu_i \neq \nu_j$
for $i \neq j$.

To produce a full solution of the Landau--Zener problem completely, 
one would have to find the 
transition amplitudes between all different states, or, in other words, 
to find the scattering matrix $S$, relating the initial state of the 
system at $t = -\infty$ and the final state at $t = +\infty$:
\begin{equation}
\psi (t = +\infty) = \hat{S} \psi (t = -\infty) \,. 
\end{equation}
The Landau-Zener theory provides an analytic solution of this problem for $n=2$.
Presently, it is not known whether a generalization of this solution exists
for $n>2$.

Here, I will pursue a less ambitious goal. 
%It turns out that for this model it is possible to find the 
%two transition probabilities: $w_{1 \to 1}$ and $w_{n \to n}$. 
I will show how one can obtain the two transition 
amplitudes $S_{11}$  and  $S_{nn}$. In our `$\nu$-ordered'
basis notation, this corresponds to the transitions
between the lowest and the highest adiabatic
energy state: $S_{1 \to 1} = S_{A1\to An}$, 
and $S_{n \to n} = S_{An \to A1}$, where $A i$ are adiabatic states
ordered by their energies (see Fig.~\ref{fig:energy evolution}). 

%% {\it probabilities}, 
%% $w_{1 \to 1} = |S_{11}|^2$, and $w_{n \to n} = |S_{nn}|^2$.
%% Since the basis is `$\nu$-ordered', these are the probabilities
%% for transition between the lowest and the highest adiabatic
%% energy state: $w_{1 \to 1} = w_{A1\to An}$, 
%% and $w_{n \to n} = w_{An \to A1}$, where $A i$ are adiabatic states
%%  ordered by their energies. 

To compute these transition probabilities, I consider the 
%% wave function 
$n$-component state 
vector $\psi(t)$ at large times $t\to \pm \infty$. 
At these times, 
the Hamiltonian is almost diagonal, and the transitions 
between different adiabatic states are negligible. 
Therefore, the state $i$'th component $\psi_i(t)$
evolves as
\begin{eqnarray}
\label{psi-asympt}
\psi_i (t\to -\infty) &=& a_i e^{-i \phi_i (t)} \,, \\
\psi_i (t\to +\infty) &=& b_i e^{-i \phi_i (t)} \nonumber \,. 
\end{eqnarray}
Let us consider the phase 
%% of the $i$'th component,
$\phi_i (t)={\rm arg}\,\psi_i(t)$. 
In the zero-order approximation, when one completely neglects 
the off-diagonal
part $\hat{\Delta}$, the phase is entirely due to $\nu_i$:
\begin{equation}
\phi_i^{(0)} (t) =  \frac{\nu_i t^2}{2} \,. 
\end{equation}
To find the phase $\phi_i$ more accurately, one notes that
in the adiabatic approximation 
this phase is actually given by an integral of a `frozen eigenvalue' of
$\varepsilon_i (t)$ the Hamiltonian (\ref{eq:A+Bt}):
%% $\varepsilon_i (t)$:
\begin{equation}
\phi_i (t) = \int\limits^{t}\varepsilon_i (t')\, dt'
\,. 
\end{equation}
At large $t$, $\nu t \gg \Delta$, 
the eigenvalue $\varepsilon_i(t)$ can be found from perturbation
theory in $\hat{\Delta}$. To the second order in $\hat \Delta$, it is given by
\begin{equation}
\varepsilon_i^{(2)} (t) = \nu_i t + \Delta_{ii} 
+ \sum_{j \neq i} \frac{|\Delta_{ij}|^2}{(\nu_i - \nu_j) t}
\,. 
\end{equation}
Therefore, at large $t$, the phase $\phi_i(t)$ can expressed as
%% , up to the second order in $\Delta$: 
\begin{equation}
\label{phi-found}
\phi_i (t) = \frac{\nu_i t^2}{2} + \Delta_{ii} t 
+ \alpha_i \ln t + O(\hat \Delta^3)\,, 
\end{equation}
where 
\begin{equation}
\label{alpha}
\alpha_i =  \sum_{j \neq i} \frac{|\Delta_{ij}|^2}{\nu_i - \nu_j}
\,. 
\end{equation}
The key observation underlying our approach is that
the terms retained in the expansion (\ref{phi-found}) 
are the only terms that grow
indefinitely at $t \to \infty$. The higher-order terms in $\hat{\Delta}$,
written symbolically
as $O(\hat \Delta^3)$, remain finite at large $t$.

The asymptotic form of $\psi_i(t)$, 
defined by the expressions (\ref{psi-asympt}), (\ref{phi-found}) and (\ref{alpha}),  
allows one to compute the 
transition amplitudes $S_{11}$ and $S_{nn}$ as follows~\cite{LL-3-problem}.
%% The method I describe below is essentially the same as used in
%% \cite{LL-3-problem}. 
Let us consider the solution~(\ref{psi-asympt}) for 
arbitrary complex times $t$,
allowing the phase variable $\phi_i(t)$ to be complex. 
Suppose that initially the system is in $k$'th state, so 
that $a_i = 0$ for all $i \neq k$, and $a_k = 1$. 
At $t \to +\infty$, the system is, in general, described by 
all $b_i \neq 0$. One can try to continue the solution at
$t \to +\infty$ to $t \to -\infty$ along one of the two 
large half-circles, 
$t = R e^{i\theta}$:\\ 
{\it (a)} in the upper half-plane, $\theta \in [0, \pi]$, 
and \\
{\it (b)} in the lower half-plane, $\theta \in [0, -\pi]$.\\
The contour radius $R$ is assumed to be very large. 

%%%%%%%%%%%%%%%%%%% figure 2 %%%%%%%%%%%%%%%%%%%%%%%%%%%%%%%%%%%
\begin{figure}[t]
\centerline{
\begin{minipage}[t]{3.5in}
\vspace{-10pt}
\hspace{-10pt}
\centering
\includegraphics[width=3in]{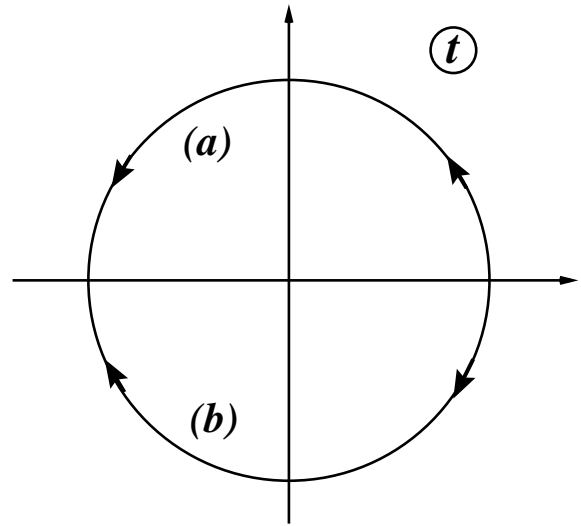}
\end{minipage}
}
\vspace{0.15cm}
\caption[]{
The contours in the complex time plane used for analytic continuation
and asymptotics matching of the 
solution for the $i=1,n$ components.
}
\label{fig:contours}
\end{figure}
%%%%%%%%%%%%%%%%%%%%%%%%%%%%%%%%%%%%%%%%%%%%%%%%%%%%%%%%%%%%%%%%%

When the state $\psi(t)$ is continued analytically along such a contour,
its different components will behave differently: some of them 
increasing, while others decreasing as a function of $|t|$. 
Indeed, the leading 
term in $\phi_i (t)$ gives for the $i$'th component 
\begin{equation}
\psi_i \propto \exp\left(\frac{\nu_i R^2}{2} \sin 2\theta + \frac{i \nu_i R^2}{2}
\cos 2\theta\right)
\,. 
\end{equation}
%
%% Applying the same reasoning as in~\cite{LL-3,LL-3-problem}, 
In matching the asymptotic expansions only 
the most rapidly increasing terms the wave function
must be retained. (For a discussion of this issue in depth, 
see~\cite{LL-3,LL-3-problem}.) For the contour (a)
the leading term
is given by the $i = n$ component, while for (b) it is the $i=1$ component.
Because of that, only $n$'th and $1$'st components of the state $\psi(t\to -\infty)$
can be determined in this way. 
Without loss of generality, I will consider the $i = 1$ case,
i.e., perform continuation along the contour {\it (b)}.

The wave function after continuing along the contour 
can be found by changing the argument of $t$ as 
\be
t \to t e^{-i\pi}
\ee
in Eq.~(\ref{phi-found}). The first and the second term contribute
only to the phase of $\psi_i(t)$, while the third term changes its modulus:
\begin{equation}
\psi_1^{(b)} (t\to -\infty) = b_1 e^{-\pi \alpha_1}  
e^{- i \frac12 \nu_1 t^2 - i \Delta_{11} t 
- i \alpha_1 \ln t }\,. 
\end{equation}
On the other hand, we assume that this state evolved from the state 
which initially had only one nonzero component, 
$|\psi_i(t)|=1$, $\psi_{j\neq i}=0$.
The time dependence of this component, $\psi_i(t)\propto e^{-i\phi_i(t)}$, with
unit modulus and
the phase given Eq.\,(\ref{phi-found}), yields
\begin{equation}
\psi_1 (t \to -\infty) = \exp\left(-{\textstyle \frac{i}2} \nu_1 t^2 - i \Delta_{11} t 
- i \alpha_1 \ln t\right)
\,. 
\end{equation}
Comparing these two expressions, we obtain
\begin{equation}
b_1 = \exp \left(\pi \alpha_1\right)
\,, 
\end{equation}
which gives the transition amplitude of the form
\begin{equation}
\label{result-1}
S_{1 \to 1} = b_1 = \exp (\pi \alpha_1) 
= \exp \left( -\pi \sum_{j \neq 1} 
            \frac{|\Delta_{i1}|^2}{\nu_j - \nu_1}\right)
\,. 
\end{equation}
Similarly, for the $i=n$ component, after going through analytical continuation over the contour {\it (a)}, we obtain
\begin{equation}
\label{result-n}
S_{n \to n} = \exp ( - \pi \alpha_n )  = 
\exp\left(
    -\pi \sum_{j \neq n} 
                \frac{|\Delta_{in}|^2}{\nu_n - \nu_j}
\right)
\,. 
\end{equation}
The resulting expressions for the transition amplitude, 
Eqs.\,(\ref{result-1}),(\ref{result-n}), bear
similarity with the Landau-Zener formula and, in fact, coincide with it 
in the $n=2$ case. Just like in the LZ problem, the dependence on the 
coupling matrix $\hat \Delta$ and on the 'rapidity' $\hat \nu$ is such that
$S\to 0$ in the adiabatic limit of slow evolution (small $\hat \nu$),
and  $S\to 1$ when the evolution is fast (large $\hat \nu$).

Eqs.~(\ref{result-1}) and~(\ref{result-n}) 
coincide with the transition amplitude found by
Brundobler and Elser~\cite{Elser93} 
in the sequential pairwise LZ level crossing approximation,
and thus prove their conjecture.

The existence of an analytic result for two transition
amplitudes may indicate that the entire 
$S$-matrix could be obtained analytically. 
It does not seem, however, that a trivial generalization 
of the above approach will be sufficient for that. 
We were not able to perform analytic continuation keeping track 
of the subleading terms, which probably indicates that this 
is not an optimal method of treating such a problem.

It is of course not unconceivable, in principle, 
that only two transition amplitudes out of total $n^2$
can be obtained in a closed form, while others cannot.
However, to the best of the author's knowledge, there are no 
other examples known where
only a subset of the $S$-matrix is calculable, 
and thus such a possibility does not seem likely. 
The author hopes that this paper will stimulate 
further work towards a more complete understanding
of multilevel Landau-Zener transitions.

\end{multicols}

\begin{thebibliography}{9}
\bibitem{LL-3-problem} We note that our method follows the line of the argument
used to study the quantum mechanical problem
of transmission through a 1D parabolic barrier
(\cite{LL-3}, \S50, Problem 4).
\end{thebibliography}
\end{document}